\documentclass[]{spie}

\title{\Large\bf Entanglement Criteria - Quantum and
Topological}

\author{Louis H. Kauffman\supit{a} and Samuel J. Lomonaco Jr.2\supit{b}
\skiplinehalf
\supit{a} Department of Mathematics, Statistics and Computer Science  
(m/c 249), 851 South Morgan Street, University of Illinois at Chicago,
Chicago, Illinois 60607-7045, USA \\
\supit{b} Department of Computer Science and Electrical Engineering, University of
Maryland Baltimore County, 1000 Hilltop Circle, Baltimore, MD 21250, USA}
 
\authorinfo{Further author information: L.H.K.  E-mail: kauffman@uic.edu,
S.J.L. Jr.: E-mail: lomonaco@umbc.edu}
 
\begin{document} 
  \maketitle 

\begin{abstract}  
This paper gives a criterion for detecting the entanglement of a quantum state, and uses it to study
the  relationship between topological and quantum entanglement.  It is
fundamental to view topological entanglements such as braids as {\em entanglement operators} and
to associate to them unitary operators that are capable of creating quantum entanglement. The entanglement
criterion is used to explore this connection.
\end{abstract}

\keywords{quantum entanglement,topological entanglement, braiding, linking, Yang-Baxter operator, non-locality}

\section{INTRODUCTION}

This paper discusses relationships between topological
entanglement and quantum entanglement. The present paper is a 
sequel to our previous work \cite{TEQE} and it explores more deeply the ideas in that earlier paper.
We propose that it is fundamental to view topological entanglements such as braids as {\em entanglement
operators} and to associate to them unitary operators that perform quantum
entanglement. One can compare the way the unitary operator corresponding to
an elementary braid has (or has not) the capacity to entangle quantum states.
One can examine the capacity of the same operator to detect
linking. The detection of linking involves working with closed braids or with
link diagrams. In both cases, the algorithms for computing link invariants are
very interesting to examine in the light of quantum computing. These algorithms
can usually be decomposed into one part that is a straight composition of unitary
operators, and hence can be seen as a sequence of quantum computer instructions, and another part that can
be seen either as preparation/detection, or as a quantum network with cycles in the underlying graph.
For the background on knotting, linking, Yang-Baxter equation and state sum invariants of links, we refer the reader
to our paper \cite{TEQE}.
\bigbreak

The paper is organized as follows.  Section 2 introduces our result 
giving a set of equations that characterize 
entanglement of a quantum state. This section then shows how this criterion can be used to analyze a general class of solutions
to the Yang-Baxter equation and their corresponding link invariants. The section ends with a discussion of 
invariants of local unitary transformations in relation to this algebraic criterion for entanglement.
Section 3 is a discussion of 
the structure of entanglement in relation to measurement. In particular, we discuss the $EPR$ thought experiment and 
discuss the Bell inequalities in the $CHSH$ formulation. We use our criterion for entanglement to show how 
unentangled states cannot violate the Bell inequalties and we give an example of an entangled state that also does not 
violate the Bell inequalities for a given choice of operators. The upshot of this discussion is that just as there is a
complex relationship between quantum entanglement and topological entanglement, there is also a complex relationship
between quantum entanglement  and non-locality. One would hope for a deeper connection between topology and non-locality.
It is our hope that this study will help in that goal.  
\bigbreak

\section{ENTANGLEMENT CRITERIA}
Let $$\phi =  \Sigma a_{\alpha} |\alpha>$$ 

\noindent 
where $\alpha$ runs over all binary strings of length
$n.$ Thus we regard $\phi$ as a general element of $V^{\otimes n}$ where $V$ is a complex vector space of
dimension two with basis $\{ |0>, |1> \}.$
\bigbreak

With $\alpha$ a binary string as above, let $|\alpha|$ denote the number of ones in the string.
Thus $|1101| = 3.$ Let $e_{i}$ denote the string of length $n$ with all zeroes except for a $1$ in the
$i$-th place. We shall write $i \in \alpha$ to mean that the $i$-th place in the string $\alpha$ is
occupied by a $1.$ Thus $i \in e_{i}$ and $2 \in 11.$ 
\bigbreak

\noindent {\bf Theorem.} The state $\phi =  \Sigma a_{\alpha} |\alpha>$ is unentangled if and only if
the following equations are satisfied for each coefficient in $\phi.$:
$$a_{0 \cdots 0}^{|\alpha| -1}a_{\alpha} = \Pi_{i \in \alpha}a_{e_{i}}.$$
\bigbreak

\noindent {\bf Proof.} If $\phi$ is unentangled then $\phi$ has the form of an $n$-fold tensor 
product as shown below, with $k$ a complex constant 
\vspace{5mm}

$$\phi = k(|0\cdots0> + A_{1\cdots0}|1\cdots0>)(|0\cdots0> + A_{01\cdots0}|01\cdots0>)$$
$$\cdots(|0\cdots0> + A_{0\cdots1}|0\cdots1>)$$
$$ = k\Pi_{i=1}^{n}(|e_0> + A_{e_i}|e_i>)$$
$$=k\Sigma_{\alpha}A_{\alpha}|\alpha>$$ where $\alpha$ runs over all binary strings of length $n$,
$A_{0\cdots0} = A_{e_0} = 1,$ and $$A_{\alpha} = \Pi_{i \in \alpha}A_{e_{i}}.$$ Since here
$$a_{\alpha} = kA_{\alpha},$$ it follows at once from this decomposition that 
$$a_{0 \cdots 0}^{|\alpha| -1}a_{\alpha} = \Pi_{i \in \alpha}a_{e_{i}}.$$ 
\smallbreak
\noindent Conversely, if the coefficients of $\phi$ satisfy this formula, then it is easy to see that
$\phi$ factorizes in the above form. This completes the proof of the Theorem.
\bigbreak

\noindent{\bf Remark.} The simplest example of this theorem is the case of two entangled qubits.
By the theorem, $$\phi = a_{00}|00> + a_{01}|01> + a_{10}|10> + a_{11}|11>$$ is unentangled exactly
when $$a_{00}a_{11} = a_{10}a_{01}.$$ This is the criterion that we checked by hand earlier in the paper.
For three qubits we have the equations
$$a_{000}a_{110} = a_{100}a_{010}$$
$$a_{000}a_{101} = a_{100}a_{001}$$
$$a_{000}a_{011} = a_{010}a_{001}$$
$$a_{000}^{2}a_{111} = a_{100}a_{010}a_{001}.$$
\bigbreak

We are now in a position to compare topological and quantum entanglement
for the larger class of solutions to the Yang-Baxter equation that we 
mentioned in our previous paper \cite{TEQE}. Recall that if $M_{\alpha, \beta}$ is a matrix with entries on the 
unit circle ($\alpha$ and $\beta$ range over all binary strings of length $n$), then we can define
$$R|\alpha, \beta> = M_{\alpha, \beta}|\beta, \alpha>,$$ and $R$ is a unitary solution to the Yang-Baxter equation.
See \cite{TEQE} for our previous discussion of this solution. 
\bigbreak

Corresponding to a link diagram $K$, we define \cite{TEQE} a 
state summation $S_{K}$ by summing over all assignments of binary strings $\alpha$ to each component of the link $K,$
(colorings of the diagram $K$)
and taking the product of the matrix entries $M_{\alpha, \beta}$ associated via $R$ to each crossing in the colored 
diagram. It then easy to see that if $K$ is a link of two components $K_1$ and $K_2$, then 

$$S_{K} = \Sigma_{\alpha \ne \beta} M_{\alpha,\alpha}^{w(K_1)} M_{\beta, \beta}^{w(K_2)}
M_{\alpha,\beta}^{2lk(K_1,K_2)}  + \Sigma_{\alpha} M_{\alpha,\alpha}^{w(K_1)} M_{\alpha,
\alpha}^{w(K_2)} M_{\alpha,\alpha}^{2lk(K_1,K_2)}.$$ Here $w(K_i)$ denotes the writhe of the component $K_i.$
That is, $w(K_i)$ denotes the sum of the signs of all the self-crossings of $K_i.$ Recall that the linking number
of $K$, denoted $lk(K) = lk(K_1,K_2),$ is one-half the sum of the signs of the crossings shared by $K_1$ and $K_2.$
Finally, the writhe of $K$, denoted $w(K),$ stands for the sum of all the signs in the diagram $K$ whether they are between
two components, or with a component and itself. The following formula is an immediate consequence of these definitions
$$w(K) = w(K_1) + w(K_2) + 2lk(K_1,K_2).$$

\noindent In order to separate out the topological dependence so that we can see how this state summation
can detect the linking number of the link, it is useful to assume that $M_{\alpha,\alpha} = \lambda$ is a constant
independent of the binary string $\alpha.$ We shall make this assumption from now on. We can then write the formula 
for the state sum in the form
$$S_{K} = \Sigma_{\alpha \ne \beta} \lambda^{w(K_1)} \lambda^{w(K_2)}
M_{\alpha,\beta}^{2lk(K_1,K_2)}  + \Sigma_{\alpha} \lambda^{w(K_1)} \lambda^{w(K_2)} \lambda^{2lk(K_1,K_2)}$$
$$= \Sigma_{\alpha \ne \beta} \lambda^{w(K)} 
(M_{\alpha,\beta}^2 /\lambda^2)^{lk(K_1,K_2)}  + \Sigma_{\alpha} \lambda^{w(K)}$$
$$= \lambda^{w(K)} \Sigma_{\alpha \ne \beta} 
(M_{\alpha,\beta}^2 /\lambda^2)^{lk(K_1,K_2)}  + 2^{n}.$$ Thus we obtain the topological invariant $Z_{K}$ defined by 
the equation
$$Z_{K} = \lambda^{-w(K)}S_{K} = \Sigma_{\alpha \ne \beta} 
(M_{\alpha,\beta}^2/\lambda^2)^{lk(K_1,K_2)}  + 2^{n}.$$ We conclude, as in the case of two qubits, that $Z_K$ can detect
linking number so long as $M_{\alpha, \beta}^2 \ne \lambda^2.$
\bigbreak

Now lets return the the matrix $R$ and see about its entanglement capabilities. We are assuming that all the $M_{\alpha,
\alpha}$ are equal to $\lambda.$ Then if $\phi = \Sigma_{\alpha, \beta}|\alpha, \beta>,$ then 
$$R\phi = \Sigma_{\alpha, \beta}M_{\beta, \alpha}|\alpha, \beta>.$$ Using our entanglement criterion
(and writing $0$ for the zero string $0\cdots0$), we conclude that
the state $R\phi$ is unentangled exactly when the following equations are satisfied for all $\alpha$ and $\beta.$
$$\lambda^{|\alpha| + |\beta| - 1}M_{\alpha, \beta} =  \Pi_{i \in \alpha}M_{e_{i},0}\Pi_{j \in \beta}M_{0,e_j}.$$ In the
case $\alpha = \beta$ this equation becomes
$$\lambda^{|\alpha| + |\alpha| - 1}M_{\alpha, \alpha} =  \Pi_{i \in \alpha}M_{e_{i},0}\Pi_{j \in \alpha}M_{0,e_j}$$
$$\lambda^{|\alpha| + |\alpha| - 1}\lambda =  \Pi_{i \in \alpha}M_{e_{i},0}\Pi_{j \in \alpha}M_{0,e_j}$$
$$\lambda^{2|\alpha|} =  \Pi_{i \in \alpha}M_{e_{i},0}\Pi_{j \in \alpha}M_{0,e_j}$$ Thus, letting
$$m_{\alpha, 0} = \Pi_{i \in \alpha}M_{e_{i},0}$$ and
$$m_{0,\alpha} = \Pi_{j \in \alpha}M_{0,e_j}$$  we have
$$\lambda^{2|\alpha|} = m_{\alpha,0} m_{0,\alpha}$$ and
$$\lambda^{|\alpha| + |\beta| - 1}M_{\alpha, \beta} =  m_{\alpha,0} m_{0,\beta}.$$ From these formulas we find 
that $$m_{0,\alpha}m_{\alpha,0}m_{0,\beta} m_{\beta,0}\lambda^{-2}M_{\alpha, \beta}^{2} =
m_{\alpha,0}m_{0,\beta}m_{\alpha,0}m_{0,\beta}.$$ Hence
$$m_{0,\alpha}m_{\beta,0}\lambda^{-2}M_{\alpha, \beta}^{2} =
m_{0,\beta}m_{\alpha,0}.$$ Therefore
$$M_{\alpha, \beta}/\lambda^{2} =
(m_{\alpha,0}/m_{0,\alpha})(m_{0,\beta}/m_{\beta,0}).$$

\noindent The state $R\phi$ is unentangled exactly when this last equation is satisfied. We see from this that 
{\it if the matrix $M$ is symmetric, i.e. if $M_{\alpha,\beta} = M_{\beta,\alpha}$ for all $\alpha$ and $\beta$}
then the invariant $Z_{K}$ detects linking exactly when $R\phi$ is an entangled state.  On the other hand,
if $M$ is not symmetric, then the invariant can detect linking even when the state $R\phi$ is unentangled. This 
is the generalization of our previous results, using the entanglement criteria proved here. The generalization 
shows that while there is no necessary relation between quantum entanglement and the ability to detect
topological linking, there are cases of invariants where the two properties are directly related to one another.

\subsection{More About Entanglement Criteria}
An element of the unitary group $U(2)$ can be represented by a matrix $U$ of the type shown below, with
$\lambda$ and $\mu$ complex numbers such that $\lambda \bar{\lambda} + \mu \bar{\mu} = e^{i\theta}$ so that 
$Det(U) = e^{i\theta}.$  
$$U =  \left(
\begin{array}{cc}
     \lambda & \mu \\
     -\bar{\mu}  & \bar{\lambda}
\end{array}\right) 
$$

\noindent We wish to consider the relationship between our algebraic criteria for entanglement and the results of
performing local unitary transformations on a state. To this end note that 

$$U|0> = \lambda|0> - \bar{\mu}|1>$$

$$U|1> = \mu|0> + \bar{\lambda}|1>,$$ and that if 

$$\psi = \Sigma a_{\alpha 0 \beta} |\alpha 0 \beta> +  a_{\alpha 1 \beta}
|\alpha 1 \beta>, $$ then 

$$(I^k \otimes U \otimes I^l)\psi =  \Sigma a_{\alpha 0 \beta} |\alpha>
(\lambda|0> - \bar{\mu}|1>)| \beta> +
a_{\alpha 1 \beta} |\alpha>(\mu|0> + \bar{\lambda}|1>)|\beta>
$$

$$ = \Sigma (\lambda a_{\alpha 0 \beta}   + \mu a_{\alpha 1 \beta} )
|\alpha 0 \beta>  + 
( -\bar{\mu} a_{\alpha 0 \beta}  +\bar{\lambda}a_{\alpha 1 \beta} )
|\alpha 1 \beta>
 $$ 
 
 $$ = \Sigma a^{'}_{\alpha 0 \beta} |\alpha 0 \beta>  + a^{'}_{\alpha 1
\beta}|\alpha 1 \beta>.
 $$

\noindent Thus
 
 $$  a^{'}_{\alpha 0 \beta} = \lambda a_{\alpha 0 \beta}   + \mu a_{\alpha
1 \beta}$$
 
 $$ a^{'}_{\alpha 1 \beta} = -\bar{\mu} a_{\alpha 0 \beta}
+\bar{\lambda}a_{\alpha 1 \beta}.$$
  
  Let
  \smallbreak
  
  $$v_{\alpha i \beta}=\left(
\begin{array}{cc}
     a_{\alpha 0 \beta} \\
     a_{\alpha 1 \beta}
\end{array}
\right)$$
  
$$v'_{\alpha i \beta}=\left(
\begin{array}{cc}
     a^{'}_{\alpha 0 \beta} \\
     a^{'}_{\alpha 1 \beta}
\end{array}
\right)$$

Then

$$v'_{\alpha i \beta} = Uv_{\alpha i \beta}$$

Hence if

$$M =  \left(
\begin{array}{cc}
       a_{\alpha 0 \beta} & a_{\gamma 0 \delta}\\
     a_{\alpha 1 \beta} & a_{\gamma 1 \delta}
\end{array}\right) 
$$

\noindent then $|Det(M)|^2$ is invariant under the local coordinate
transformation $U$  since  with 

$$M' =  \left(
\begin{array}{cc}
       a'_{\alpha 0 \beta} & a'_{\gamma 0 \delta}\\
     a'_{\alpha 1 \beta} & a'_{\gamma 1 \delta}
\end{array}\right) 
$$

\noindent we have that $M' = UM.$
\smallbreak

There are many choices of these two by two determinants that are invariant
under local coordinate transformations.
It is easy to show that they all vanish for an unentangled state.
We conjecture that if all of them vanish, then the state is
unentangled. This point will be taken up in a sequel to this paper.
{\it Note that the value of $|Det(M)|^2$ is only an invariant for the specific local transformation with 
which it is associated}, but that in the case of two qubits the non-zero values of $|Det(M)|^2$ are exactly
$|a_{00}a_{11} - a_{01}a_{10}|^2$ which we know to determine entanglement in this case. We see here that for two qubits
the value of $|Det(M)|^2$ {\em is} invariant under all local unitary transformations of the state. This can also be
verified by  a density matrix calculation. 
\smallbreak

\section{A REMARK ABOUT $EPR$, ENTANGLEMENT AND BELL'S INEQUALITY} 
It is remarkable that the simple algebraic situation
of an element in a tensor product that is not itself a a tensor product of
elements of the factors corresponds to subtle nonlocality in physics. It helps to
place this algebraic structure in the context of a gedanken experiment to see
where the physics comes in. Consider $$S = (|0>|1> + |1>|0>)/\sqrt{2}.$$

\noindent
In an $EPR$ thought experiment, we think of two ``parts" of this state that are
separated in space.  We want a notation for these parts and suggest the following:

$$L = (\{|0>\}|1> + \{|1>\}|0>)/\sqrt{2},$$

$$R = (|0>\{|1>\} + |1>\{|0>\})/\sqrt{2}.$$

\noindent In the left
state  $L$, an observer can only observe the left hand factor. In the right state $R$,
an observer can only observe the right hand factor.
These ``states"  $L$ and $R$ together comprise the $EPR$ state $S,$ but they are
accessible individually just as are the two photons in the usual thought
experiement.  One can transport $L$ and $R$ individually and we shall write

$$S = L*R$$

\noindent to denote that they are the ``parts"  (but not tensor factors) of $S.$
\bigbreak

The curious thing about this formalism is that it includes a little bit of
macroscopic physics implicitly, and so it makes it a bit more apparent what $EPR$
were concerned about.  After all, lots of things that we can do to $L$ or $R$ do not
affect $S.$ For example, transporting $L$ from one place to another, as in the
original experiment where the photons separate.  On the other hand, if Alice has
$L$ and Bob has $R$ and Alice performs a local unitary transformation on ``her" tensor
factor, this applies to both $L$ and $R$ since the transformation is actually being
applied to the state $S.$ This is also a ``spooky action at a distance" whose
consequence does not appear until a measurement is made.
\bigbreak

To go a bit deeper it is worthwhile seeing what entanglement, in the sense of tensor indecomposability,
has to do with the structure of the $EPR$ thought experiment. To this end, we look at the structure of
the Bell inequalities using the $CHSH$ formalism as explained in book by Nielsen and Chuang \cite{NielsenChuang}. For this
we use the  following observables with eigenvalues $\pm 1.$
$$Q =  \left(
\begin{array}{cc}
       1 & 0\\
       0 & -1
\end{array}\right)_1, 
$$
$$R =  \left(
\begin{array}{cc}
       0 & 1\\
       1 & 0
\end{array}\right)_1, 
$$
$$S =  \left(
\begin{array}{cc}
       -1 & -1\\
       -1 & 1
\end{array}\right)_2/\sqrt{2}, 
$$
$$T =  \left(
\begin{array}{cc}
       1 & -1\\
       -1 & -1
\end{array}\right)_2/\sqrt{2}. 
$$ 

\noindent The subscripts $1$ and $2$ on these matrices indicate that they are to operate on the first and 
second tensor factors, repsectively, of a quantum state of the form
$$\phi = a|00> + b|01> + c|10> + d|11>.$$ To simplify the results of this calculation we shall here assume that 
the coefficients $a,b,c,d$ are real numbers. We calculate the quantity 
$$\Delta = <\phi|QS|\phi> + <\phi|RS|\phi> + <\phi|RT|\phi> - <\phi|QT|\phi>,$$ finding that
$$\Delta = (2 - 4(a+d)^2 + 4(ad -bc))/\sqrt{2}.$$ Classical probability calculation with random variables of value
$\pm 1$ gives the value of $QS + RS + RT - QT = \pm 2$ (with each of $Q$, $R$, $S$ and $T$ equal to $\pm 1$).
Hence the classical expectation satisfies the Bell inequality
$$E(QS) + E(RS) + E(RT) - E(QT) \le 2.$$ That quantum expectation is not classical is embodied in the fact that
$\Delta$ can be greater than $2.$ The classic case is that of the Bell state
$$\phi = (|01> - |10>)/\sqrt{2}.$$ Here $$\Delta =  6/\sqrt{2} > 2.$$ In general we see that the following inequality is 
needed in order to violate the Bell inequality
$$(2 - 4(a+d)^2 + 4(ad -bc))/\sqrt{2} > 2.$$ This is equivalent to 
$$(\sqrt{2}-1)/2 < (ad-bc) - (a+d)^2.$$ Since we know that $\phi$ is entangled exactly when $ad-bc$ is non-zero, this
shows that an unentangled state cannot violate the Bell inequality. This formula {\em also} shows that it is possible
for a state to be entangled and yet not violate the Bell inequality. For example, if
$$\phi = (|00> - |01> + |10> + |11>)/2,$$ then $\Delta(\phi)$ satisfies Bell's inequality, but $\phi$ is an entangled 
state. We see from this calculation that entanglement in the sense of tensor indecomposability, and entanglement in the
sense of Bell inequality violation for a given choice of Bell operators are not equivalent concepts. On the other hand,
Benjamin Schumacher has pointed out \cite{S} that any entangled two-qubit state will violate Bell inequalities for an
appropriate choice of operators. We will expand the
discussion of this point in a joint paper \cite{Us} under preparation. This
deepens the context for our question of the relationship between topological entanglement and quantum entanglement. The
Bell inequality violation is an indication of true quantum mechanical entanglement. One's intuition suggests that it is
{\em this} sort of entanglement that should have a topological context. We will continue in the search for that context. 
\bigbreak 

\section*{ACKNOWLEDGMENTS}  
Most of this effort was sponsored by the Defense
Advanced Research Projects Agency (DARPA) and Air Force Research Laboratory, Air
Force Materiel Command, USAF, under agreement F30602-01-2-05022. Some of this
effort was also sponsored by the National Institute for Standards and Technology
(NIST). The U.S. Government is authorized to reproduce and distribute reprints
for Government purposes notwithstanding any copyright annotations thereon. The
views and conclusions contained herein are those of the authors and should not be
interpreted as necessarily representing the official policies or endorsements,
either expressed or implied, of the Defense Advanced Research Projects Agency,
the Air Force Research Laboratory, or the U.S. Government. (Copyright 2003.) It
gives the first author great pleasure to thank Fernando Souza and Heather Dye for
conversations in the course of preparing this paper. \bigbreak

 \end{document}